\RequirePackage{snapshot}
\pdfoutput=1

\documentclass[apl,amsmath,amssymb,amnsfonts,twocolumn,a4paper]{revtex4-1}
\usepackage[T1]{fontenc}
\usepackage{geometry}
\usepackage{graphicx}
\usepackage{color}

\usepackage{setspace}
\usepackage{booktabs}

\begin{document}

\title{Landau-Lifshitz-Bloch equation for exchange coupled grains} %Title of paper

\author{Christoph Vogler}
\email{christoph.vogler@tuwien.ac.at}
\affiliation{Institute of Solid State Physics, Vienna University of Technology, Wiedner Hauptstrasse 8-10, 1040 Vienna, Austria}

\author{Claas Abert}
\author{Florian Bruckner}
\author{Dieter Suess}
\affiliation{Christian Doppler Laboratory for Advanced Magnetic Sensing and Materials, Institute for Solid State Physics, Vienna University of Technology, Wiedner Hauptstrasse 8-10, 1040 Vienna, Austria}

\date{\today}

\begin{abstract}
Heat assisted recording is a promising technique to further increase the storage density in hard disks. Multilayer recording grains with graded Curie temperature is discussed to further assist the write process. Describing the correct magnetization dynamics of these grains, from room temperature to far above the Curie point, during a write process is required for the calculation of bit error rates. We present a coarse grained approach based on the Landau-Lifshitz-Bloch (LLB) equation to model exchange coupled grains with low computational effort. The required temperature dependent material properties such as the zero-field equilibrium magnetization as well as the parallel and normal susceptibilities are obtained by atomistic Landau-Lifshitz-Gilbert (LLG) simulations. Each grain is described with one magnetization vector. In order to mimic the atomistic exchange interaction between the grains a special treatment of the exchange field in the coarse grained approach is presented.
\end{abstract}

                             % Classification Scheme.
%Use showkeys class option if keyword
                              %display desired
\maketitle %\maketitle must follow title, authors, abstract and \pacs

Over the last decades the areal storage density of hard disk drives (HDD) continuously increased \cite{wood_future_2009}. In order to keep up this increase many inventions on both sides the magnetic write head and the recording medium were necessary. One of the first improvements beyond the pure scaling of all involved parts of a HDD was the introduction of anisotropic magneto-resistive write heads. A significant increase in the sensitivity of magnetic heads became possible due to the discovery of the effect of giant magneto-resistance \cite{baibich_giant_1988,gruenberg_magnetic_1989}, which is the basis for all modern magnetic read heads. Concerning the recording medium, where all information is written and stored, the invention of antiferromagnetic coupled media and especially the transition from longitudinal to perpendicular recording \cite{iwasaki_co-cr_1978,iwasaki_perpendicular_1980} have to be mentioned. A recent improvement \cite{suess_multilayer_2006} uses recording grains consisting of many different materials with graded anisotropy instead of grains with single phases. Nevertheless the areal storage density increase of HDD slowed down with the state-of-the-art recording techniques, because with decreasing particle size, magnetically harder recording grains have to be used in order that the stored information remains thermally stable. In principle the magnetic field required to write a graded media grain can be arbitrarily reduced with enough layers. But it is technically not possible to produce such grains with a continuous change of their anisotropy constant.\newline
Heat assisted recording could be the next step to provide a further increase in the areal storage density of HDD. In this technique the recording medium is locally heated near or above the Curie temperature ${T_\mathrm{C}}$ to be able to reverse the magnetic moments of recording grains with very high coercivity, like FePt. In combination with an additional write assistance of grains with graded Curie temperatures the further continuous increase of the areal storage density of HDD is trusted for the next years.\newline
There are several ways to handle the effect of temperature in micromagnetism. The most common way to account for thermal fluctuations, acting on the magnetic moments of a ferromagnetic particle is to incorporate a random thermal field in the equation of motion. At zero temperature the  integration of the Landau-Lifshitz-Gilbert (LLG) equation is an established method to describe magnetization dynamics. The problem of the LLG equation is that the magnetization length is proposed to be fixed, independent of temperature. Once the spatial discretization of a ferromagnetic particle is not atomistic this restriction is wrong at high temperatures, because the phase transition from the ferromagnetic state to the paramagnetic state at the Curie temperature can not be modeled in each macroscopic simulation cell. It is well know from molecular field theory that the length of the total magnetization of an ensemble of magnetic moments decreases with increasing temperature and finally ends up with zero length at ${T_\mathrm{C}}$. Hence a powerful high temperature micromagnetic equation should reproduce this behavior in each computational cell. The Landau-Lifshitz-Bloch (LLB) equation derived by Garanin~\cite{garanin_fokker-planck_1997} fulfills the requirement and the absolute value of the magnetization is no longer a constant. It links between the LLG equation at low temperatures and the Bloch equation at high temperatures. Since in an atomistic LLG model each atom of a magnetic particle has to be described with one spin the simulation already becomes computationally very expensive for small grains with lateral dimensions of a few nanometers. In contrast, the LLB equation permits to compute large areas of a particle, or even the whole grain, with just one magnetic moment. Hence, LLB simulations are very fast compared to their atomistic LLG counterparts. As a result bit error rates of magnetic grains with realistic dimensions can be computed, which would not be possible with atomistic LLG calculations.\newline
The structure of the paper is as follows: In Sec.~\ref{sec:models} the basic equations of motion of both used models, the atomistic LLG model and the LLB model are introduced. Section~\ref{sec:input_LLB_from_VAMPIRE} shows how temperature dependent material functions, which are required by the LLB model, are obtained by atomistic LLG simulations using VAMPIRE~\cite{evans_atomistic_2013}. Since the presented coarse grained LLB model is used to describe heat assisted recording for graded Curie temperature grains we derive an intergrain exchange expression from the Heisenberg Hamiltonian in Sec.~\ref{sec:intergrain_exchange}. Finally the results of the model for a realistic high/low ${T_\mathrm{C}}$ recording grain subject to a temperature pulse are presented in Sec.~\ref{sec:results_comparison}. Additionally a comparison to the according atomistic LLG results is given in this section.
\section{Dynamic equations}
\label{sec:models}
As already mentioned the Landau-Lifshitz-Bloch (LLB) equation describes the magnetization dynamics of magnetic particles at high temperatures without the restriction of a fixed magnetization length and thus allows for its longitudinal relaxation. The validity of the LLB was already proven in various publications~\cite{garanin_thermal_2004,chubykalo-fesenko_dynamic_2006,atxitia_micromagnetic_2007,kazantseva_towards_2008,chubykalo-fesenko_dynamic_2006,schieback_temperature_2009,bunce_laser-induced_2010,evans_stochastic_2012,mcdaniel_application_2012,greaves_magnetization_2012,mendil_speed_2013}. In its most recent formulation~\cite{evans_stochastic_2012} it has the following form
\begin{eqnarray}
\label{eq:LLB}
  \frac{d \boldsymbol{m}}{dt}= &-&\mu_0{\gamma'}\left( \boldsymbol{m}\times \boldsymbol{H}_{\mathrm{eff}}\right) \nonumber \\
  &-&\frac{\alpha_\perp\mu_0 {\gamma'}}{m^2} \left \{ \boldsymbol{m}\times \left [ \boldsymbol{m}\times \left (\boldsymbol{H}_{\mathrm{eff}}+\boldsymbol{\xi}_{\perp}  \right ) \right ] \right \}\nonumber \\
  &+&\frac{\alpha_\parallel  \mu_0{\gamma'}}{m^2}\boldsymbol{m}\left (\boldsymbol{m}\cdot\boldsymbol{H}_{\mathrm{eff}}  \right )+\boldsymbol{\xi}_{\parallel},
\end{eqnarray}
where $\gamma'$ is the reduced electron gyromagnetic ratio ($\gamma'=|\gamma_{\mathrm{e}}|/(1+\lambda^2)$ with $|\gamma_{\mathrm{e}}|=1.760859708\cdot10^{11}$\,(Ts)$^{-1}$), $\mu_0$ is the vacuum permeability and $\boldsymbol{m}$ is the reduced magnetization $\boldsymbol{M}/M_0$, with the saturation magnetization at zero temperature $M_0$. In addition $\alpha_\parallel$ and $\alpha_\perp$ are dimensionless temperature dependent longitudinal and transverse damping parameters given by
\begin{equation}
 \alpha_\perp=\begin{cases}\lambda\left( 1-\frac{T}{T_{\mathrm{C}}} \right) & T<T_{\mathrm{C}}\\ \alpha_\parallel & T\geq T_{\mathrm{C}}\end{cases},\quad\alpha_\parallel=\lambda \frac{2T}{3T_{\mathrm{C}}}.
\end{equation}
The coupling of the spin to the heat bath on an atomistic level is described by $\lambda$. $T_{\mathrm{C}}$ donates the Curie temperature. The longitudinal and perpendicular thermal fields are denoted with $\eta=\parallel$ and $\eta=\perp$ respectively. $\boldsymbol{\xi}_{\eta}$ consist of white noise random numbers with zero mean and a variance of 
\begin{equation}
  \left \langle \xi_{\eta,i}(t,\boldsymbol{r})\xi_{\eta,j}({t}',\boldsymbol{r}') \right \rangle = 2D_\eta \delta_{ij}\delta(\boldsymbol{r}-\boldsymbol{r}')\delta(t-{t}'),
\end{equation}
where the diffusion constants $D_\eta$ follow from the fluctuation-dissipation theorem to
\begin{eqnarray}
\label{eq:thermalField_strangths}
  D_\perp&=&\frac{\left (\alpha_\perp-\alpha_\parallel  \right )k_{\mathrm{B}} T}{ \gamma' \mu^2_0 M_0 V \alpha^2_\perp}\nonumber\\
  D_\parallel&=&\frac{\alpha_\parallel \gamma' k_{\mathrm{B}} T}{M_0 V}.
\end{eqnarray}
The effective magnetic field $\boldsymbol{H}_{\mathrm{eff}}$ in Eq.~\ref{eq:LLB} consists of four contributions in our model. Beside the external field $\boldsymbol{H}_{\mathrm{ext}}$ it contains the intergrain exchange field $\boldsymbol{H}_{\mathrm{iex}}$, which is discussed in more detail in Sec.~\ref{sec:intergrain_exchange}, the anisotropy field $\boldsymbol{H}_{\mathrm{ani}}$ and the internal exchange field $\boldsymbol{H}_{\mathrm{J}}$
\begin{equation}
 \boldsymbol{H}_{\mathrm{eff}}=\boldsymbol{H}_{\mathrm{ext}}+\boldsymbol{H}_{\mathrm{iex}}+\boldsymbol{H}_{\mathrm{ani}}+\boldsymbol{H}_{\mathrm{J}}.
\end{equation}
The anisotropy field has the following compact form
\begin{equation}
  \label{eq:Hani}
   \boldsymbol{H}_\mathrm{ani}=\frac{1}{\widetilde{\chi}_{\perp}}\left( m_x\boldsymbol{e}_{x}+m_y\boldsymbol{e}_{y}\right),
\end{equation}
with the perpendicular susceptibility $\widetilde{\chi}_{\perp}$. Here it is assumed that the easy axis, arising from the uniaxial anisotropy of the crystal structure, points along the $z$-direction. Since the anisotropy constant $K_1$ and the magnetization $M$ are both temperature dependent, $\widetilde{\chi}_{\perp}$ is also a function of temperature. Two further temperature dependent material functions appear in the internal exchange field $\boldsymbol{H}_{\mathrm{J}}$ controlling the length of the magnetization, which is defined as
\begin{equation}
\label{eq:blochField}
 \boldsymbol{H}_{\mathrm{J}}=\begin{cases} \frac{1}{2\widetilde{\chi}_{\parallel}}\left( 1-\frac{m^2}{m^2_{\mathrm{e}}} \right)\boldsymbol{m} & T\lesssim T_{\mathrm{C}}\\ -\frac{1}{\widetilde{\chi}_{\parallel}} \left( 1+\frac{3}{5}\frac{T_{\mathrm{C}}}{T-T_{\mathrm{C}}}m^2 \right)\boldsymbol{m}& T\gtrsim T_{\mathrm{C}}\end{cases}.
\end{equation}
In this equation $m_{\mathrm{e}}$ is the zero field reduced equilibrium magnetization. The perpendicular and longitudinal susceptibilities are specified as
\begin{equation}
 \label{eq:susceptibilityParallel}
  \widetilde{\chi}_{\eta}=\left( \frac{d{m_\eta}}{d{H}_{\mathrm{ext},\eta}} \right)_{{H}_{\mathrm{ext},\eta} \rightarrow 0}.
\end{equation}
To integrate the LLB equation at arbitrary temperatures the detailed temperature dependence of $m_{\mathrm{e}}$, $\widetilde{\chi}_{\parallel}$ and $\widetilde{\chi}_{\perp}$ has to be known. LLG simulations with an atomistic spatial discretization of the underlying ferromagnetic particle as well as a mean field ansatz can be used for this purpose.\newline
For the atomistic approach we use the LLG code VAMPIRE~\cite{evans_atomistic_2013}, where the dynamic equation of motion is implemented as follows
\begin{eqnarray}
 \label{eq:atomisticLLG}
 \frac{d\boldsymbol{S}_k}{dt}=&-&\mu_0 \gamma'\left \{\boldsymbol{S}_k \times\left ( \boldsymbol{H}_{\mathrm{eff},k}+\boldsymbol{\xi}_k \right )   \right \}\nonumber \\
 &-&\mu_0 \gamma' \lambda \left \{ \boldsymbol{S}_k \times \left [ \boldsymbol{S}_k \times \left (  \boldsymbol{H}_{\mathrm{eff},k}+\boldsymbol{\xi}_k\right ) \right ] \right \}.
\end{eqnarray}
Here $\boldsymbol{S}_k$ is a unit vector denoting the direction of the spin of lattice site $k$. The random thermal field again has white noise properties with zero mean and a variance of
\begin{equation}
 \label{eq:variance_LLG}
 \left \langle \xi_{i,k}(t)\xi_{j,l}({t}') \right \rangle = \frac{2\lambda k_{\mathrm{B}}T}{\gamma \mu_{\mathrm{S}}\mu_0^2} \delta_{ij}\delta_{k,l}\delta(t-{t}'),
\end{equation}
where $i,j$ are the Cartesian components of the thermal field and $k,l$ represent the lattice sites. The effective magnetic field $\boldsymbol{H}_{\mathrm{eff},k}$ acting on spin $k$ can be expressed as the derivative of the spin Hamiltonian with respect to $\boldsymbol{S}_k$ 
\begin{equation}
 \boldsymbol{H}_{\mathrm{eff},k}=-\frac{1}{\mu_{\mathrm{S}}\mu_0}\frac{\partial \mathcal{H}}{\partial\boldsymbol{S}_k},
\end{equation}
with $\mu_{\mathrm{S}}$ being the atomistic magnetic moment. VAMPIRE uses a typical spin Hamiltonian containing exchange energy, uniaxial anisotropy energy and Zeemann energy as follows
\begin{eqnarray}
 \label{eq:Heisenberg}
 \mathcal{H}=-\sum_{k,l}J_{k,l}\boldsymbol{S}_k\boldsymbol{S}_l&-&K_1\sum_k S_{z,i}^2\nonumber \\
 &-&\mu_{\mathrm{S}}\sum_k \boldsymbol{H}_{\mathrm{ext}} \cdot \boldsymbol{S}_k.
\end{eqnarray}
Beside the geometry of the particle, the Heisenberg exchange parameters $J_{k,l}$, the uniaxial anisotropy constant $K_1$ and the atomistic spin moment $\mu_{\mathrm{S}}$ are the main input parameters in this model.\newline
Compared to the atomistic LLG equation (Eq.~\ref{eq:atomisticLLG}) the LLB equation has two additional contributions, namely the last two terms on the right-hand side of Eq.~\ref{eq:LLB}, describing the changes in the length of the magnetization with temperature and ensuring that even the magnetization of a particle, represented with just one spin, vanishes at the Curie temperature.
\section{Temperature dependent material functions}
\label{sec:input_LLB_from_VAMPIRE}
For the solution of the LLB equation the temperature dependence of the zero field equilibrium magnetization $m_\mathrm{e}(T)$, the parallel susceptibility $\widetilde{\chi}_{\parallel}(T)$ and the normal susceptibility $\widetilde{\chi}_{\perp}(T)$ are required. These informations are obtained by atomistic simulations using VAMPIRE. In this paper we use cylindrical layers with 5\,nm height and a basal plane with a diameter of 5\,nm. We model two different materials, a hard magnetic (HM) with the material parameters of FePt and a soft magnetic (SM), Fe like one. Since it was reported \cite{suess_prediction_2014} that high damping in the soft magnetic part of grains with graded Curie temperature improves their recording properties a damping constant of 1.0 is used for the SM material. For simplicity both materials are assumed to have a simple cubic crystal structure.
\begin{table}[h!]
  \centering
  \vspace{0.5cm}
  \begin{tabular}{c c c}
    \toprule
    \toprule
      & HM & SM \\
    \midrule
    $K_1$\,[J/m$^3$] & $6.6\cdot 10^6$ & 0.0 \\
    $J_{k,l}$\,[J/link] & $5.18\cdot 10^{-21}$ & $7.18\cdot 10^{-21}$ \\
    $\mu_{\mathrm{S}}$\,[$\mu_{\mathrm{B}}$] & 1.7 & 1.7 \\
    $J_{\mathrm{S}}$\,[T] & 1.43 & 1.43 \\
    $a$\,[nm] & 0.24 & 0.24 \\
    $\lambda$ & 0.1 & 1.0\\
    $T_{\mathrm{C}}$\,[K] & 536.47 & 820.78\\
    \bottomrule
    \bottomrule
  \end{tabular}
  \caption{\small (color online) Simulation input parameters of a hard magnetic (HM), FePt like material and a soft magnetic (SM), Fe like material. $K_1$ is the anisotropy constant, $J_{k,l}$ is the Heisenberg exchange parameter in Joule per interaction link, $\mu_{\mathrm{S}}$ is the atomistic magnetic moment in units of the Bohr magneton $\mu_{\mathrm{B}}$, $J_{\mathrm{S}}$ is the corresponding saturation magnetization in the LLB model, $a$ is the lattice constant of the used simple cubic lattice, $\lambda$ is the damping constant and $T_{\mathrm{C}}$ is the fitted Curie temperature.}
  \label{tab:mat_match}
\end{table}
The detailed parameters are illustrated in Tab.~\ref{tab:mat_match}. For any other system the procedure works similarly.
\subsection{Calculation of $m_\mathrm{e}(T)$}
\label{Calculation of me}
We simulate 100 trajectories of 20000 time steps with an integration step of $10^{-15}$\,s (after 20000 equilibration steps) for each temperature value in the range of $0-800$\,K for the HM material using VAMPIRE. 
\begin{figure}[!h]
\includegraphics{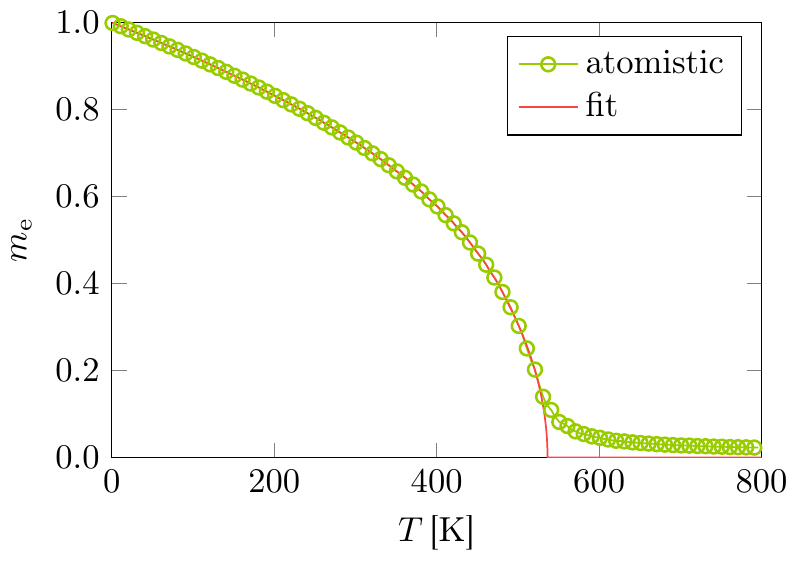}
  \caption{\small (color online) Zero field equilibrium magnetization $m_{\mathrm{e}}$ versus temperature, calculated from an atomistic model of the HM material (see Tab.~\ref{tab:mat_match}). The red solid line shows a fit, representing an infinite system.}
  \label{fig:me_fit_hard_lambda0p1}
\end{figure}
Figure~\ref{fig:me_fit_hard_lambda0p1} illustrates the atomistic result for $m_\mathrm{e}(T)$ after averaging over the 100 calculated trajectories. The plot clearly shows finite size effects. The LLB equation requires temperature dependent functions for an infinite system, because the Curie point has to be properly defined. Thus the atomistic data are fitted with true critical behavior near $T_{\mathrm{C}}$. A fit with $f(T)=\frac{C}{T-T_{\mathrm{C}}}$, where $C$ is the Curie constant, extrapolates to $T_{\mathrm{C}}=536.47$\,K. The same procedure is used to calculate $m_{\mathrm{e}}$ for the SM layer.
\subsection{Calculation of $\widetilde{\chi}_{\parallel}(T)$ and $\widetilde{\chi}_{\perp}(T)$}
\label{Calculation of chi}
According to the spin fluctuation model the transverse and parallel susceptibilities can be obtained by the fluctuations of the magnetization components between sequent time steps in the atomistic model as follows
\begin{equation}
\label{eq:chi_eta}
 \widetilde{\chi}_\eta=\frac{\mu_{\mathrm{S}}N}{k_{\mathrm{B}}T}\left( \left \langle m_\eta^2 \right \rangle - \left \langle m_\eta \right \rangle^ 2 \right ).
\end{equation}
Transverse and parallel denote directions with respect to the easy axis of the investigated particle. To be consistent with Eq.~\ref{eq:susceptibilityParallel}, where the susceptibilities are defined with respect to an external applied field, both the preferred magnetic direction and the direction of the magnetic field are assumed to be parallel. In Eq.~\ref{eq:chi_eta} $N$ is the number of spins, $T$ is the temperature and
\begin{equation}
 m_\eta=\frac{1}{N}\sum_{i=k}^{N}S_{\eta,k},
\end{equation}
is the average magnetization along the direction of $\eta$. All fluctuations are calculated at zero applied field. 
\begin{figure}[!h]
\includegraphics{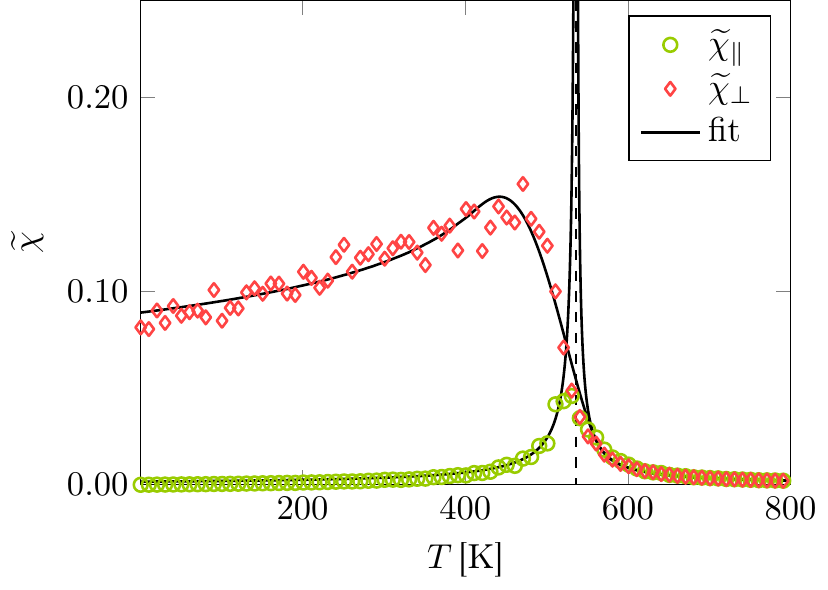}
  \caption{\small (color online) Transverse and parallel susceptibilities versus temperature of a HM material (see Tab.~\ref{tab:mat_match}), obtained by the fluctuations of the magnetization components in an atomistic model. The lines show fit functions extrapolating to the critical behavior of an infinite system. The dashed line indicates $T_{\mathrm{C}}$.}
  \label{fig:chi_fit_hard_lambda0p1}
\end{figure}
In the case of the HM material the corresponding fluctuations, obtained by 100 atomistic trajectories of 20000 time steps ($\Delta t=10^{-15}$\,s, after 20000 equilibration steps) at each temperature in the range of $0-800$\,K, are illustrated in Fig.~\ref{fig:chi_fit_hard_lambda0p1}. The expected critical behavior of $\widetilde{\chi}_\parallel$ at the Curie point can be clearly seen. Above $T_{\mathrm{C}}$ the particle is paramagnetic and thus the susceptibilities in all directions become equal. As already mentioned in the last section the LLB equation requires temperature dependent functions for an infinite system. From the spin fluctuation model it is known that the longitudinal susceptibility is proportional to $1/(T-T_{\mathrm{C}})$ around the Curie temperature, which is used as fit function.\newline
At low temperatures
\begin{equation}
\label{eq:chi_perp_lowT}
 \widetilde{\chi}_\perp=\frac{M_0 m_{\mathrm{e}}^2}{2K_1(T)}
\end{equation}
holds. If $K_1(T)$ is expressed with a power law $K_1(T) \propto m_{\mathrm{e}}^{c}$ the final piecewise fit functions for the susceptibilities are as follows
\begin{equation}
\label{eq:fit_parallel}
 \widetilde{\chi}_\parallel=\begin{cases}
 \frac{c_1}{T_\mathrm{C}-T} & T<T_\mathrm{C} \\ 
 \frac{c_2}{T-T_\mathrm{C}} & T>T_\mathrm{C}
\end{cases}
\end{equation}
\begin{equation}
 \widetilde{\chi}_\perp=\begin{cases}
 c_3 m_{\mathrm{e}}^{c_4} & T<<T_\mathrm{C} \\
 \frac{c_5}{T-T_\mathrm{C}} & T>T_\mathrm{C}
\end{cases},
\end{equation}
where $c_1-c_5$ are the fit parameters which have to be determined for the investigated particle. In the intermediate temperature range where $\widetilde{\chi}_\perp$ is still undefined a fourth order polynomial is used, which is continuously differentiable at the connection points to the low and high temperature functions. With the remaining degree of freedom of the polynomial the atomistic data are then fitted. The intersection points which delimit the parts with different fit behavior of $\widetilde{\chi}_\perp$ are chosen to minimize the mean squared displacement of the fit and the atomistic fluctuations in the whole temperature range. For the HM material with uniaxial anisotropy the illustrated procedure to compute the required temperature dependent susceptibilities is straightforward.\newline 
It is different in the case of the SM material (see Tab.~\ref{tab:mat_match}), which has small or no uniaxial anisotropy, but still strong exchange. Without external field such a particle is superparamagnetic.
\begin{figure}[!h]
\includegraphics{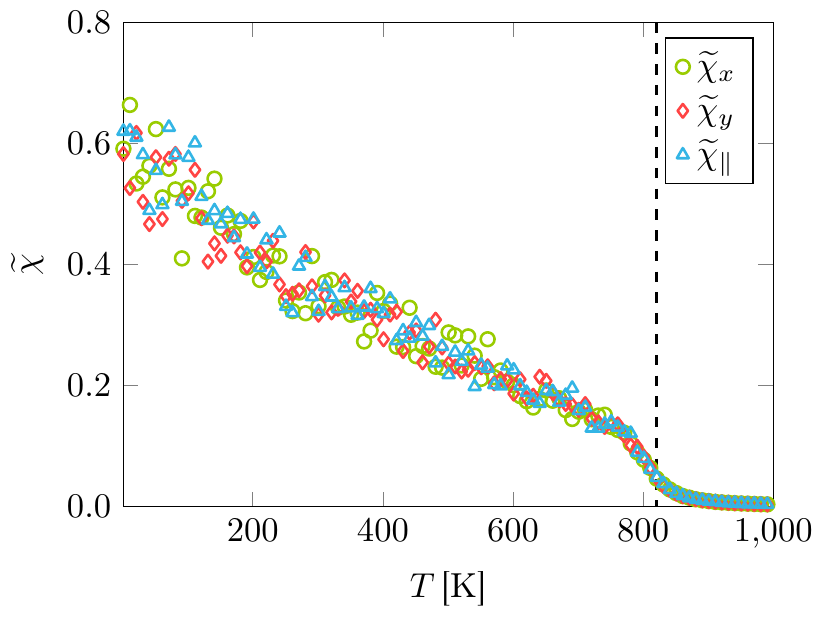}
  \caption{\small (color online) Susceptibilities versus temperature of the SM material (see Tab.~\ref{tab:mat_match}), obtained by the fluctuations of the magnetization components in an atomistic LLG model. The dashed line indicates $T_{\mathrm{C}}$. The data belong to a cylindrical particle consisting of the SM material (see Tab.~\ref{tab:mat_match}) and are simulated with VAMPIRE.}
  \label{fig:chi_fit_soft_lambda1p0}
\end{figure}
Averaging over the corresponding magnetization fluctuations of 100 trajectories at each temperature from $0-1000$\,K yields the susceptibilities shown in Fig.~\ref{fig:chi_fit_soft_lambda1p0}. All components of the susceptibility coincide, because the particle has no preferred magnetic direction. Hence there does not exist a critical behavior of $\widetilde{\chi}_\parallel$. Above the Curie point we again find the $(T-T_{\mathrm{C}})^{-1}$ dependence of all susceptibility components. For low temperatures Fig.~\ref{fig:chi_fit_soft_lambda1p0} does not reflect the full range of the magnetization fluctuations. The thermal field, which drives the magnetization, is small and hence the simulated trajectories are too short to capture the full magnitude of the susceptibilities. From a physical point of view $\widetilde{\chi}_\perp$ should be infinite, because Eq.~\ref{eq:chi_perp_lowT} holds. Furthermore a very soft magnetic ferromagnet like Fe reacts fast to an external magnetic field for temperatures below $T_{\mathrm{C}}$ and aligns its magnetization along the field. For these reasons the transverse susceptibility has to be set to infinity in the LLB model of a soft magnetic particle with no uniaxial anisotropy. \newline
In contrast $\widetilde{\chi}_\parallel$ must not be infinity, as it ensures the magnetization length in the LLB equation to remain in the vicinity of $m_{\mathrm{e}}$, according to the internal exchange field $\boldsymbol{H}_{\mathrm{J}}$ (Eq.~\ref{eq:blochField}). Since $\widetilde{\chi}_\parallel$ can not be obtained by fluctuations of the $z$-component of the magnetization, we propose to extract it from the variance in the magnetization length as 
\begin{equation}
\label{eq:chi_m}
 \widetilde{\chi}_m=b \cdot \operatorname{Var}(|\boldsymbol{m}|)=b\left( \left \langle |\boldsymbol{m}|^2 \right \rangle - \left \langle |\boldsymbol{m}| \right \rangle^ 2 \right ).
\end{equation}
\begin{figure}[!h]
\includegraphics{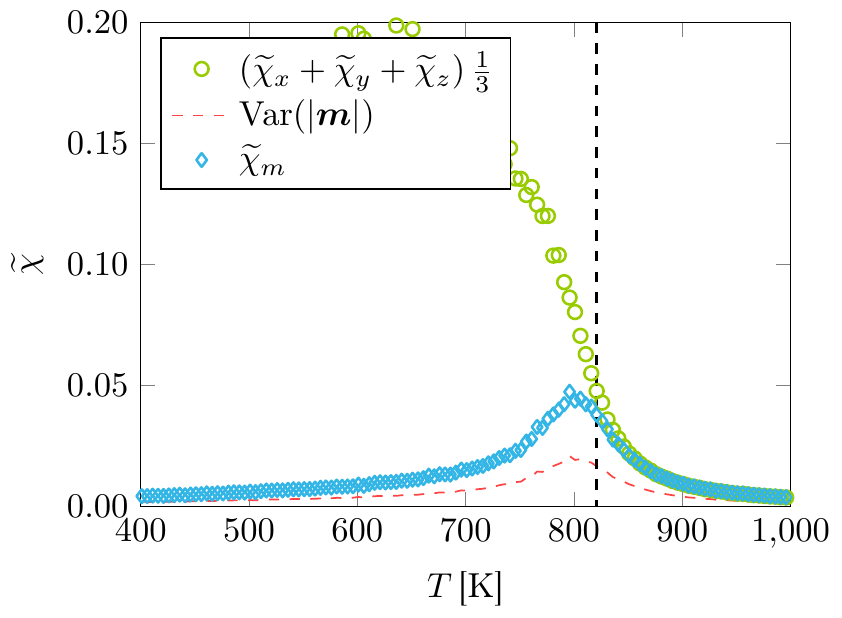}
  \caption{\small (color online) Variance of the magnetization length fitted to the average of the Cartesian components of the susceptibility, above $T_{\mathrm{C}}$. The resulting function of $\widetilde{\chi}_m$ serves as parallel susceptibility. The dashed line indicates $T_{\mathrm{C}}$. The data belong to a cylindrical particle consisting of the SM material (see Tab.~\ref{tab:mat_match}) and are simulated with VAMPIRE.}
  \label{fig:chi_fit_soft_lambda1p0_2}
\end{figure}
Figure~\ref{fig:chi_fit_soft_lambda1p0_2} displays that the fluctuations of the magnetization length are smaller than the fluctuations of its components, because the length can not change its sign. But $\operatorname{Var}(|\boldsymbol{m}|)$ shows critical behavior. With the proportionality factor $b$ in Eq.~\ref{eq:chi_m} the length fluctuations are scaled to the average fluctuations of its Cartesian components ($\sum_i\frac{1}{3} \widetilde{\chi}_i$) above the Curie point (Fig.~\ref{fig:chi_fit_soft_lambda1p0_2}).
\begin{figure}[!h]
\includegraphics{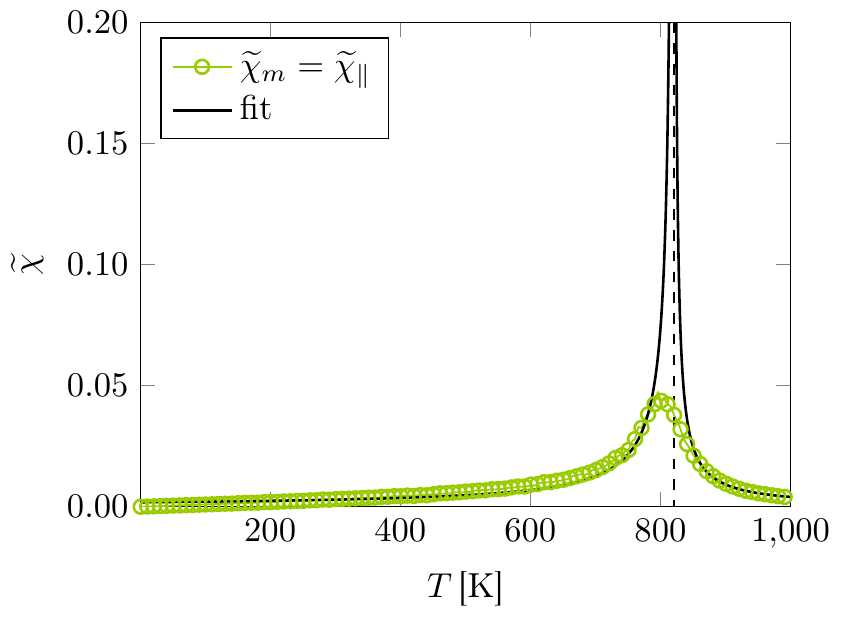}
  \caption{\small (color online) Fit of the parallel susceptibility of the SM material according to Eq.~\ref{eq:fit_parallel}. The dashed line indicates $T_{\mathrm{C}}$.}
  \label{fig:chi_fit_soft_lambda1p0_3}
\end{figure}
The fit functions listed in Eq.~\ref{eq:fit_parallel} are then applied to the resulting $\widetilde{\chi}_m$, yielding the parallel susceptibility which is needed for the LLB model, as shown in Fig.~\ref{fig:chi_fit_soft_lambda1p0_3}.\newline
\begin{figure}[!h]
\includegraphics{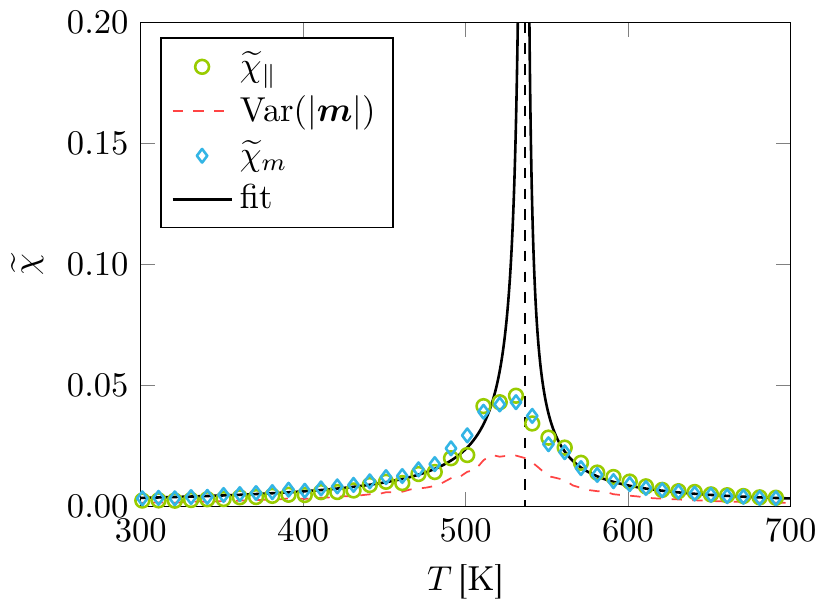}
  \caption{\small (color online) Identical fluctuations of the $z$-component of the magnetization ($\widetilde{\chi}_\parallel$) and its length ($\widetilde{\chi}_m$), after scaling the latter. The dashed line indicates $T_{\mathrm{C}}$. The data belong to a cylindrical particle consisting of the HM material (see Tab.~\ref{tab:mat_match}) and are simulated with VAMPIRE.}
  \label{fig:chi_fit_hard_lambda0p1_2}
\end{figure}
The above presented procedure to obtain the parallel susceptibility for the LLB model can in principle be applied to an arbitrary particle, ranging from very soft to very hard magnetic. Figure~\ref{fig:chi_fit_hard_lambda0p1_2} illustrates that the scaled fluctuations of the magnetization length correspond to the fluctuations of its $z$-component, also in case of the HM material.
\section{Intergrain exchange}
\label{sec:intergrain_exchange}
\begin{figure}[!h]
  \centering
  \includegraphics{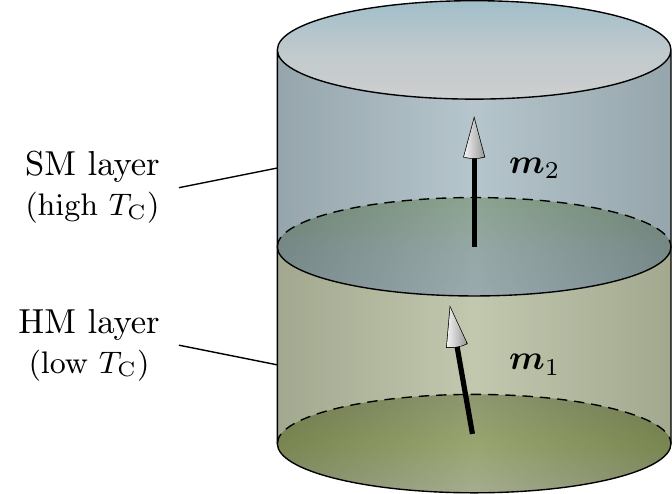}
  \caption{\small (color online) Grain model consisting of a stack of two layers with high and low Curie temperatures, coupled via an intergrain exchange interaction on the boundary surface. Each layer is represented as a single magnetization vector ($\boldsymbol{m}_1$ and $\boldsymbol{m}_2$).}
  \label{fig:layer_model}
\end{figure}
Since we aim to model high/low $T_{\mathrm{C}}$ grains we have to describe the coupling of different material layers. In the coarse grained model each layer is represented by one magnetization vector, which are coupled via an intergain exchange interaction on the boundary surface, as shown in Fig.~\ref{fig:layer_model}. In this work we restrict ourselves to two layers, but it is easy to extend the model to an arbitrary layer number. In order to derive the intergrain exchange the Heisenberg Hamiltonian, which gives the total exchange energy across the grains boundary surface, serves as a starting point
\begin{equation}
 \mathcal{H}=-\sum_{\mathrm{ss}}J_{kl}\boldsymbol{S}_k\boldsymbol{S}_l.
\end{equation}
Here ss indicates the sum over all surface spins. The exchange integrals $J_{kl}$ are assumed to be independent of the lattice site. With the unit vectors $\boldsymbol{u}_k$ and $\boldsymbol{u}_l$ along the spin directions the Hamiltonian reads
\begin{equation}
 \mathcal{H}=-JS^2\sum_{\mathrm{ss}}\boldsymbol{u}_k\boldsymbol{u}_l.
\end{equation}
In a simple cubic lattice each spin just has one nearest neighbor on the opposing side of an interface. In that case one can rewrite $\sum_{\mathrm{ss}}$ as sum over all spins on the surface of layer 1 each interacting with its neighboring spin in layer 2 
\begin{equation}
 \mathcal{H}=-2JS^2\sum_{k}\boldsymbol{u}_{k,1}\boldsymbol{u}_{k,2}.
\end{equation}
Now we perform the transition from the atomistic to the LLB description where all spins in each layer are described with just one magnetization vector. Since the magnetization length is not conserved the unit vectors can be written as
\begin{equation}
\label{eq_heisenberg_hamiltonian}
 \mathcal{H}=-2JS^2\frac{F}{a^2}\frac{\boldsymbol{m}_1}{m_1}\frac{\boldsymbol{m}_2}{m_2},
\end{equation}
where $F$ is the area of the interaction surface and $a$ the lattice constant in the atomistic model. Hence $F/a^2$ is the number of spins on the boundary surface. With Eq.~\ref{eq_heisenberg_hamiltonian} the intergrain exchange field of layer 1 can be derived by taking the derivative of the exchange energy with respect to the layer's magnetic moment
\begin{equation}
\label{eq:exchange_field_def}
 \boldsymbol{H}_{\mathrm{iex},1} = -\frac{1}{V\mu_0M_0}\frac{\partial}{\partial \boldsymbol{m}}_1\mathcal{H}.
\end{equation}
The intergrain exchange field calculates to
\begin{eqnarray}
  \boldsymbol{H}_{\mathrm{iex},1} &=& \frac{2JS^2F}{a^2V\mu_0M_0} \left ( \frac{\boldsymbol{m}_2 m_1 m_2 - \boldsymbol{m}_1 \boldsymbol{m}_2 \frac{\boldsymbol{m}_1}{m_1}m_2}{m_1^2 m_2^2} \right )\nonumber \\
  &=& \frac{2JS^2}{a^2d\mu_0M_0} \left ( \frac{\boldsymbol{m}_2 }{m_1 m_2} - \cos(\theta _{12}) \frac{\boldsymbol{m}_1}{m_1^2} \right ),
\end{eqnarray}
with the angle between the magnetic moments $\theta _{12}$ and the thickness $d$ of layer 1. Introducing the temperature dependent intergrain exchange constant $A_{\mathrm{iex}}(T)=JS^2$ the intergrain exchange field acting on the magnetization of layer 1 becomes
\begin{equation}
\label{eq:Hex1}
  \boldsymbol{H}_{\mathrm{iex},1}=\frac{2A_{\mathrm{iex}}(T)}{a^2d\mu_0M_0} \left ( \frac{\boldsymbol{m}_2 }{m_1 m_2} - \cos(\theta _{12}) \frac{\boldsymbol{m}_1}{m_1^2} \right ).
\end{equation}
The computation of the temperature dependence of $\boldsymbol{H}_{\mathrm{iex}}$ is in general less straightforward. For FePt the bulk exchange stiffness was successfully computed by determining the free energy and the width of a domain wall in the investigated material \cite{kazantseva_towards_2008,hinzke_domain_2008}. There also exists an approach where the dispersion relation of thermally excited spin waves yields the temperature dependence of the exchange coupling \cite{atxitia_multiscale_2010}. Both methods are computationally expensive and yield the same scaling behavior of $A_{\mathrm{iex}}\sim m^{\alpha}$. We try to keep the coarse grained LLB model as simple as possible and construct the temperature dependence of the intergrain exchange analytically from the according dependencies of the bulk exchange constants in the interacting layers. These are described with a power law of the magnetization length $A(T)\propto m_{\mathrm{e}}^\alpha(T)$, which holds at least for low temperatures. In many cases it is also a suitable description at high temperatures \cite{atxitia_multiscale_2010}.\newline
From a physical point of view we ask for symmetric exchange constants with equivalent $A_{12}(T)$ and $A_{21}(T)$. There are two obvious possibilities for symmetric intergrain exchange constants
\begin{itemize}
 \item an arithmetic mean of the bulk values 
\begin{equation}
 A_{\mathrm{iex}}(T)=A_{\mathrm{iex}}(0)\frac{m_{\mathrm{e},1}^\alpha(T)+m_{\mathrm{e},2}^\beta(T)}{2}
\end{equation}
 \item or an the geometric mean of the bulk values
 \begin{equation}
 A_{\mathrm{iex}}(T)=A_{\mathrm{iex}}(0)\sqrt{m_{\mathrm{e},1}^\alpha(T) m_{\mathrm{e},2}^\beta(T)}.
\end{equation}
\end{itemize}
$\alpha$ and $\beta$ are the corresponding power law exponents for the temperature dependence of the bulk exchange constants of the layers. At the Curie temperature the magnetization becomes zero, thus also the intergrain exchange should vanish. Since the geometric mean is zero as soon as one of the equilibrium magnetizations vanishes, the geometric mean is the preferred formulation. Finally the full exchange field of layer 1 is
\begin{eqnarray}
  \boldsymbol{H}_{\mathrm{iex},1} &=&\frac{2A_{\mathrm{iex}}(0)\sqrt{m_{\mathrm{e},1}^\alpha(T) m_{\mathrm{e},2}^\beta(T)}}{a^2d\mu_0M_0 m_1}  \nonumber \\
  &&\cdot\left ( \frac{\boldsymbol{m}_2 }{m_2} - \cos(\theta _{12}) \frac{\boldsymbol{m}_1}{m_1} \right ).
\end{eqnarray}
Atxitia et.~al. \cite{atxitia_multiscale_2010} investigated the power law of the exchange stiffness with numerical methods and derived the exponent of FePt analytically to 1.76, which we use for the HM layer. For a generic ferromagnet with localized magnetic moments on a simple cubic lattice and in the absence of anisotropy the exponent becomes 1.66, which is used for the SM layer.
\section{Results and discussion}
\label{sec:results_comparison}
We investigate the switching behavior of a high/low $T_{\mathrm{C}}$ grain subject to a heat pulse with Gaussian profile
\begin{equation}
\label{eq:gauss_profile}
 T(t)=\left (  T_{\mathrm{peak}}-T_{\mathrm{min}} \right )e^{-\left (\frac{t-t_{\mathrm{peak}} }{t_{\mathrm{pulse}}}  \right )^2}+T_{\mathrm{min}}.
\end{equation}
The initial temperature of the pulse $T_{\mathrm{min}}$ is set to 270\,K and $t_{\mathrm{peak}}=3t_{\mathrm{pulse}}$ is valid in all simulations. The grain has a cylindrical geometry with a basal plane diameter of 5\,nm and a total height of 10\,nm and it consists of 50\,\% HM and 50\,\% SM material as introduced in Tab.~\ref{tab:mat_match}. All calculations start with a magnetization in the positive $z$-direction. An external magnetic field assists the magnetization reversal and points in the negative $z$-direction with a tilt of 10\,\%. In the atomistic simulations with VAMPIRE a simple cubic crystal lattice with a lattice constant of $a=0.24$\,nm is assumed in all parts of the grain. We compute the switching probability of the recording grain subject to heat pulses with different durations $t_{\mathrm{pulse}}$ and peak temperatures.
\begin{figure}[!h]
\centering
\includegraphics{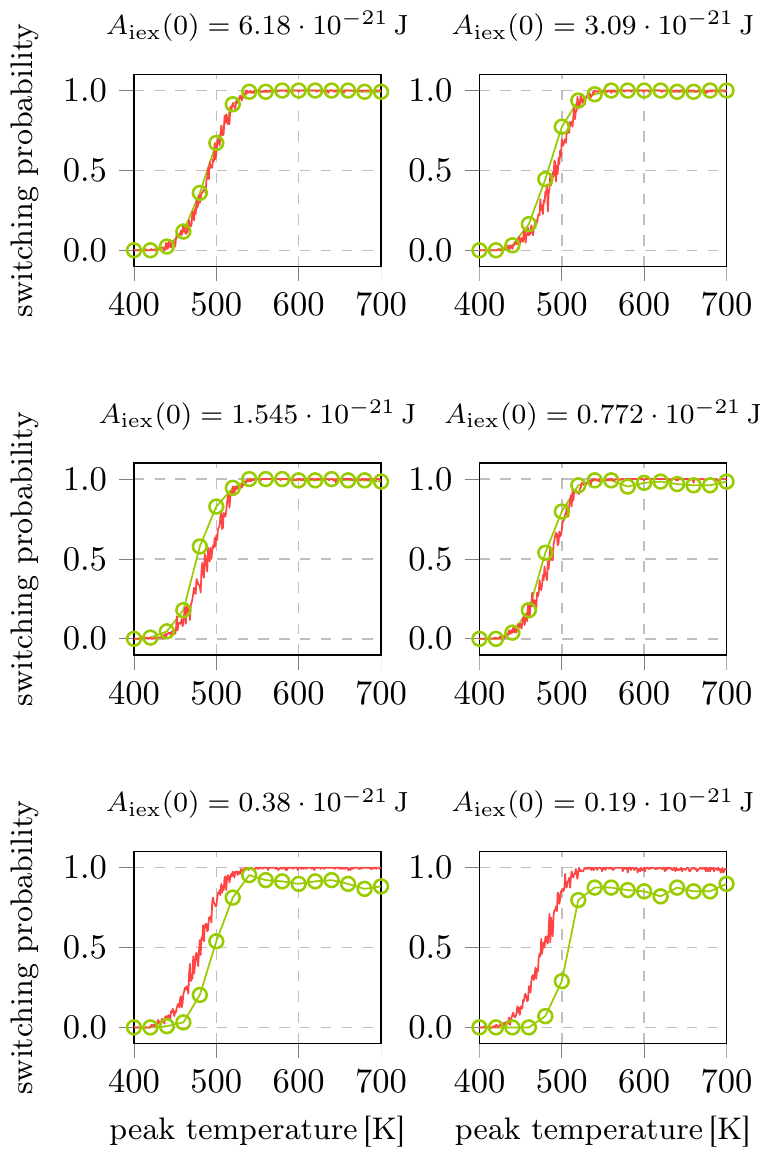}
  \caption{\small (color online) Comparison of atomistic switching probability curves (green lines with circles) with the results of the coarse grained LLB model (red solid lines) for different intergrain exchange constants ($A_{\mathrm{iex},n}(0)=6.18\cdot10^{-21}/2^n$\,J, $0\leq n\leq5$). The investigated high/low $T_{\mathrm{C}}$ grain is subject to a Gaussian heat pulse with $t_{\mathrm{pulse}}=100$\,ps and an external field with 0.5\,T strength.}
  \label{fig:llb_h0p5_100ps}
\end{figure}
Figure~\ref{fig:llb_h0p5_100ps} illustrates such switching probability curves for heat pulses with a duration of $t_{\mathrm{pulse}}=100$\,ps and an external field with $\mu_0 H_{\mathrm{ext}}=0.5$\,T. Each probability value is an average of 128 independent trajectories, computed with both the coarse grained LLB model (red solid lines) and with VAMPIRE (green lines with circles) for different intergrain exchange constants $A_{\mathrm{iex}}(0)$. The accordance is insufficient for weak intergrain exchange. Although the curves seem to agree well for strong exchange coupling, we will see in the next section that the problem is the same for large $A_{\mathrm{iex}}(0)$. Actually there exists a LLB switching probability curve which fits the VAMPIRE data much better. As long as the exchange is strong the switching probability curves do not change much and thus the agreement still seems to be well in Fig.~\ref{fig:llb_h0p5_100ps}.
\subsection{Intergrain exchange field correction}
\label{sec:correction}
In order to resolve the discrepancy between LLB and VAMPIRE simulations we examine a simpler system, consisting of two identical HM layers. We calculate the switching probabilities for 6 intergrain exchange values ($A_{\mathrm{iex},n}=5.18\cdot10^{-21}/2^n$\,J, $0\leq n\leq5$) using VAMPIRE. The applied Gaussian heat pulse has again a duration of $t_{\mathrm{pulse}}=100$\,ps and the applied field has a strength of 0.5\,T. Since the coarse grained LLB model is computationally less expensive the same probability curves for 70 values of $A_{\mathrm{iex}}(0)$ is simulated in the same range. 
\begin{figure}[!h]
\includegraphics{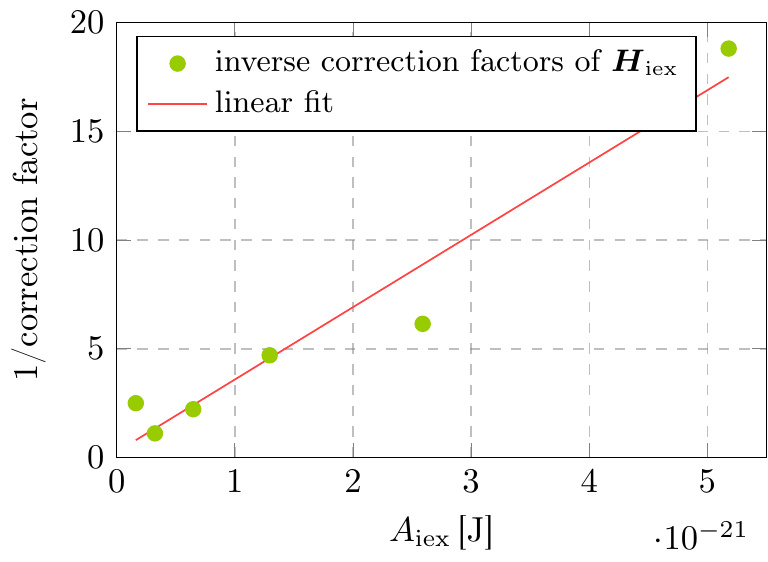}
  \caption{\small (color online) Linear fit of the inverse correction factors of the intergrain exchange field in the LLB model consisting of two HM layers (see Tab.~\ref{tab:mat_match}). The bulk exchange interaction within the layers is assumed to be $A_{\mathrm{ex}}=5.18\cdot10^{-21}$\,J.}
  \label{fig:correction}
\end{figure}
After that the LLB results are fitted to the atomistic ones, yielding correction factors for the exchange field in the LLB model as shown in Fig.~\ref{fig:correction}. For increasing intergrain exchange constant the reduction of $\boldsymbol{H}_{\mathrm{iex}}$ increases linearly. This dependence can be understood as follows: for weak coupling the exchange interaction is mainly located at the interface between the layers, but for large $A_{\mathrm{iex}}$ the domain wall is not restricted to the interface but extends towards the bulk magnets. However, in the LLB approach the grain is not discretized and the domain wall can not be formed except at the interface. Hence the domain wall energy is overestimated. For this reason a correction factor of almost $1/20$ is needed in the exchange field to reproduce the correct dynamics with the LLB model, if full bulk exchange is assumed in the calculations. It is not surprising that the correction factor nearly equals the ratio of the lattice constant and the layer thickness $a/d$, because after inserting the reduction factor in Eq.~\ref{eq:Hex1} and considering that the magnetization lengths $m_1$ and $m_2$ are almost identical in the same material, the exchange field becomes 
\begin{equation}
\label{eq:Hex_micro}
  \boldsymbol{H}_{\mathrm{iex}}=\frac{2A_{\mathrm{iex}}(T)}{ad^2\mu_0M_0 m^2} \left ( \boldsymbol{m}_2 - \cos(\theta _{12})\boldsymbol{m}_1\right ).
\end{equation}
Under the micromagnetic assumption that neighboring magnetic moments just comprise small angles $\cos(\theta _{12})\approx 1$ is valid and thus Eq.~\ref{eq:Hex_micro} becomes identical to the discretized Laplace operator (discretization length $d$) in a finite difference schema \cite{atxitia_micromagnetic_2007,kazantseva_towards_2008,schieback_temperature_2009,mendil_speed_2013}.\\
\begin{figure}[!h]
\includegraphics{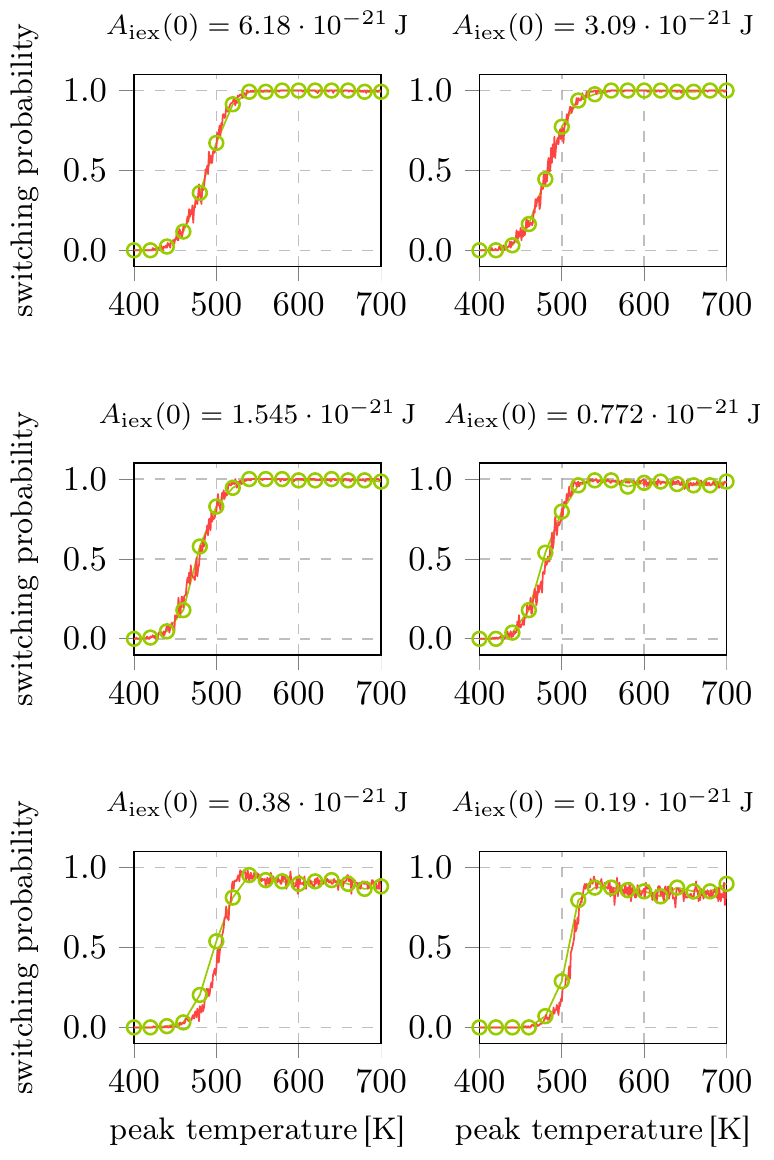}
  \caption{\small (color online) Switching probabilities as in Fig.~\ref{fig:llb_h0p5_100ps}, with a corrected intergrain exchange field at different $A_{\mathrm{iex}}(0)$. The correction function is constructed as described in Sec.~\ref{sec:correction}.}
  \label{fig:llb_h0p5_100ps_corrected}
\end{figure}
\subsection{Results with corrected $\boldsymbol{H}_{\mathrm{iex}}$}
\label{sec:corr_results}
The case is similar for graded grains consisting of layers with different bulk exchanges. The presented construction of a linear correction fit function for the exchange field from several switching probability simulations at different strengths of the intergrain exchange works well as Fig.~\ref{fig:llb_h0p5_100ps_corrected} shows. The figure again displays the switching probabilities of a high/low $T_{\mathrm{C}}$ grain with an applied heat pulse of $t_{\mathrm{pulse}}=100$\,ps and an applied field of 0.5\,T. The coarse grained LLB model with two magnetic moments produces the same switching probabilities as the atomistic model with over 14000 spins. Qualitatively the data demonstrate that in case of strong intergrain exchange the switching probability almost everywhere reaches 100\,\%. For weak coupling the switching probability decreases significantly at high peak temperatures and the edge of the probability curve shifts to higher peak temperatures.\newline
\begin{figure}[!h]
\includegraphics{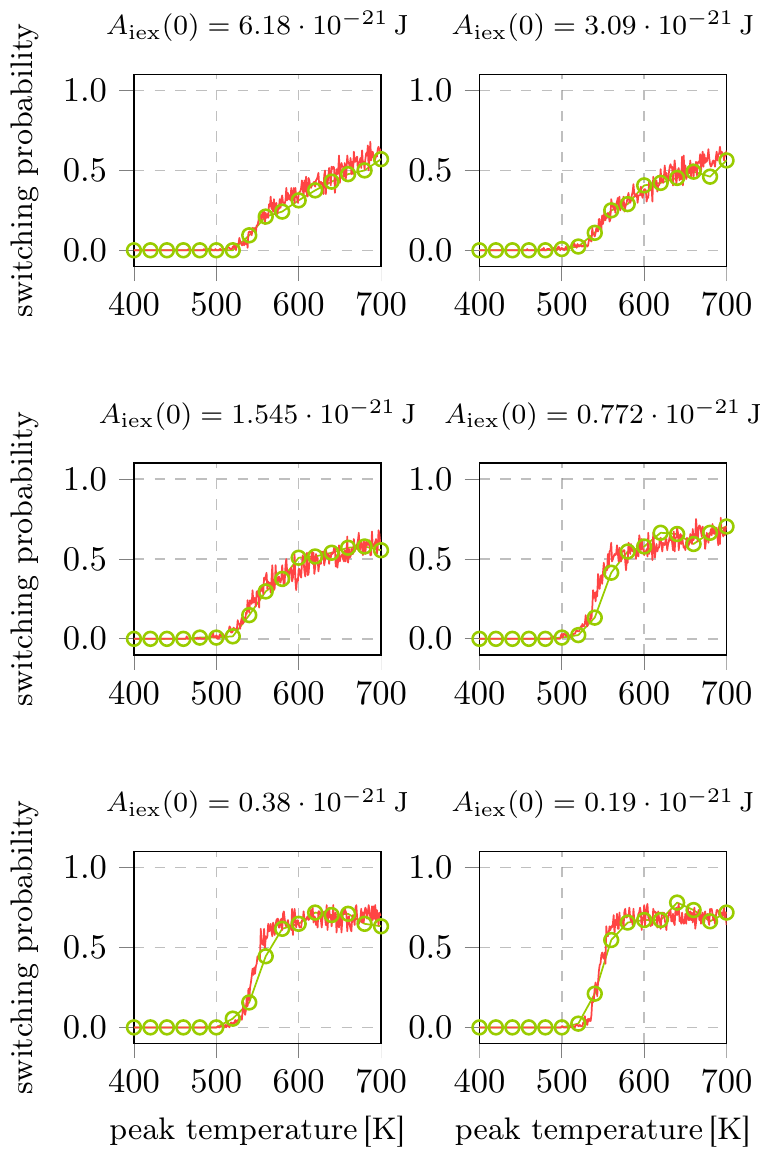}
  \caption{\small (color online) Same switching probabilities as presented in Fig.~\ref{fig:llb_h0p5_100ps} for a Gaussian heat pulse with $t_{\mathrm{pulse}}=10$\,ps. The exchange field in the LLB model is corrected as described in Sec.~\ref{sec:correction}.}
  \label{fig:llb_h0p5_10ps_corrected}
\end{figure}
The same simulations with a shorter Gaussian heat pulse with $t_{\mathrm{pulse}}=10$\,ps show again good agreement between the LLB model and atomistic simulations (Fig.~\ref{fig:llb_h0p5_10ps_corrected}). It has to be mentioned that the same correction values are used for shorter pulses, because the pulse duration does not change anything in the exchange properties at the interface.
\begin{figure}[!h]
\includegraphics{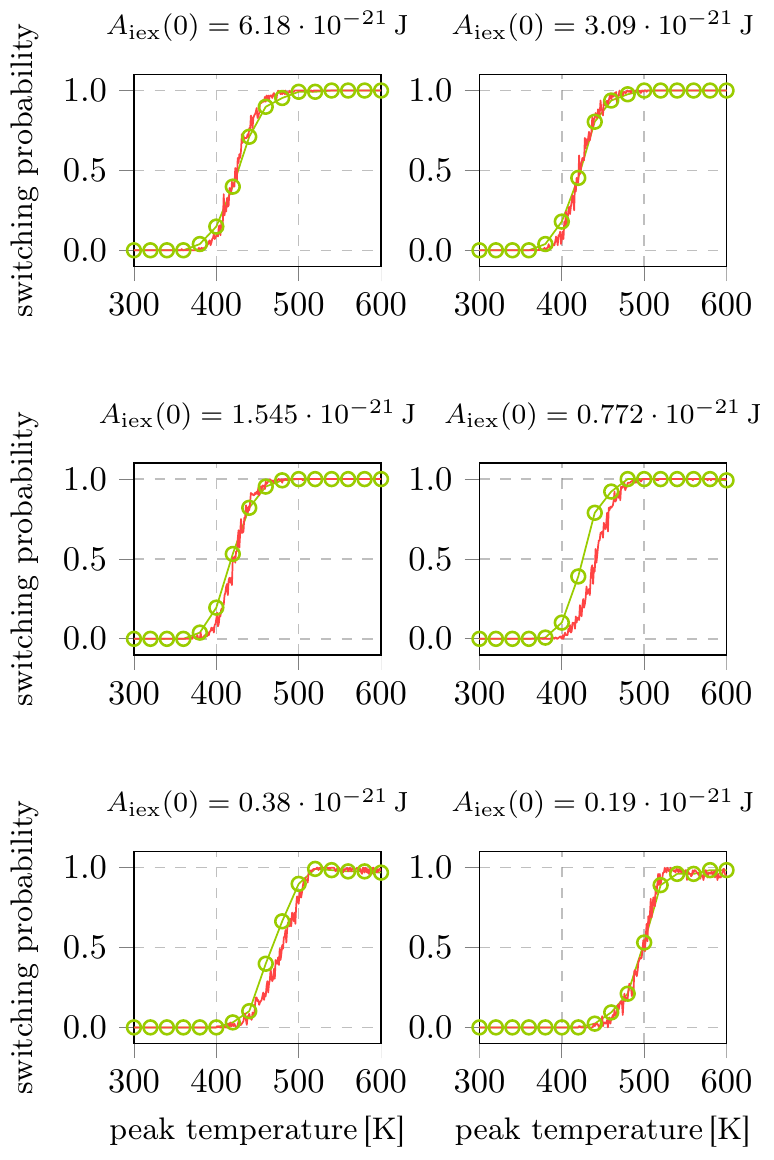}
  \caption{\small (color online) Same switching probabilities as presented in Fig.~\ref{fig:llb_h0p5_100ps} for a Gaussian heat pulse with $t_{\mathrm{pulse}}=100$\,ps and an external field with 0.8\,T strength. The exchange field in the LLB model is corrected as described in Sec.~\ref{sec:correction}.}
  \label{fig:llb_h0p8_100ps_corrected}
\end{figure}
For a stronger magnetic field the comparison of the switching probability curves is shown in Fig.~\ref{fig:llb_h0p8_100ps_corrected}. Here new corrections of the intergrain exchange field are calculated, because the field strength also influences the exchange properties at the interface. The accordance is very well except for an intermediate intergrain exchange constant of $A_{\mathrm{iex}}(0)=0.772\cdot10^{-21}$\,J. Although the edge of the switching probability curve is slightly shifted, the final switching probabilities are still correctly reproduced by the coarse grained LLB model. As expected, under the influence of a higher external field the switching probabilities remain in the vicinity of one even for weak coupling in contrast to a field of $H_{\mathrm{ext}}=0.5$\,T.\newline
%\pgfplotsset{colormap={myColMap}{[5pt] color(0pt)=(android_dark_green) color(500pt)=(white) color(1000pt)=(android_dark_red)}}
\begin{figure}[!h]
\includegraphics{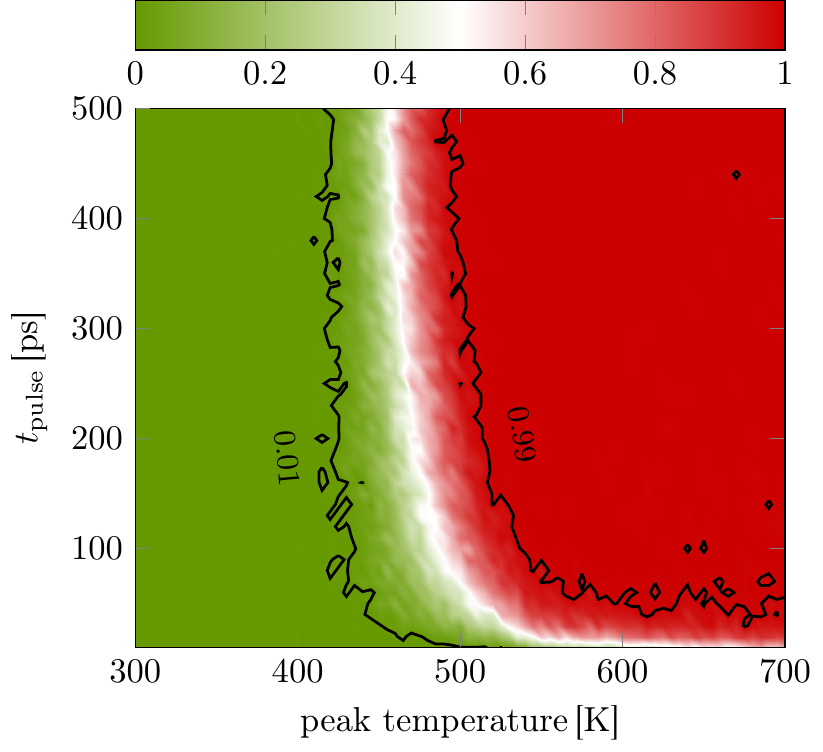}
  \caption{\small (color online) Switching probabilities of a high/low $T_{\mathrm{C}}$ recording grain subject to a Gaussian heat pulse with different lengths $t_{\mathrm{pulse}}$ and peak temperatures. The material parameters of the layers are given in Tab.~\ref{tab:mat_match} and the intergrain exchange constant at zero temperature is $A_{\mathrm{iex}}=6.18\cdot10^{-21}$\,J. Additionally an external magnetic field of 0.5\,T is applied to the grain.}
  \label{fig:phase_h0p5}
\end{figure}
\begin{figure}[!h]
\includegraphics{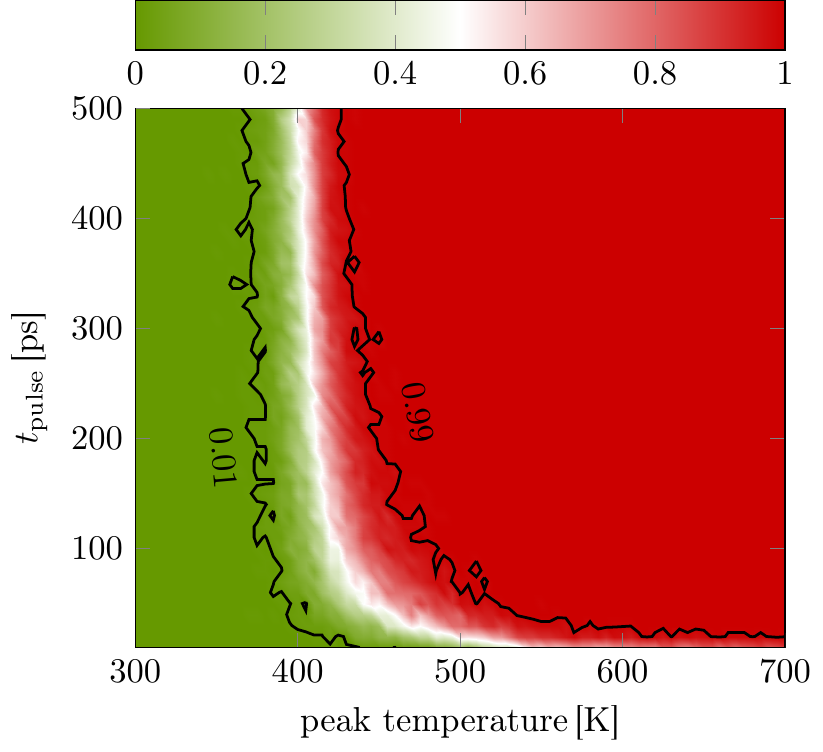}
  \caption{\small (color online) Same as Fig.~\ref{fig:phase_h0p5} with an external magnetic field of 0.8\,T.}
  \label{fig:phase_h0p8}
\end{figure}
After validating the coarse grained LLB model we can benefit from its efficiency and calculate phase diagrams of the switching probability of the high/low $T_{\mathrm{C}}$ grain for heat pulses with different durations as presented in Fig.~\ref{fig:phase_h0p5} and Fig.~\ref{fig:phase_h0p8}. Different external magnetic fields are used in the diagrams. Each of them contains a total of 4000 points. Each point shows the switching probability computed from 128 switching trajectories for different peak temperatures and pulse lengths. One trajectory with a pulse duration of $t_{\mathrm{pulse}}=100$\,ps requires almost 35 minutes of computation time on a single core machine with VAMPIRE while the same simulation finishes within 7 seconds with the LLB model. Hence the phase diagrams are difficult or even completely impossible to generate with atomistic simulations. Figures~\ref{fig:phase_h0p5} and \ref{fig:phase_h0p8} display that the switching probability does not improve much for field pulses $t_{\mathrm{pulse}}>100$\,ps and thus an optimal heat assistance which guarantees fast and reliable switching has a pulse duration of $100-150$\,ps and a peak temperature of about 600\,K for an applied field of 0.5\,T and 500\,K for $H_{\mathrm{ext}}=0.8$\,T, respectively. This is valid if an intergrain exchange at zero temperature of $A_{\mathrm{iex}}(0)=6.18\cdot10^{-21}$\,J is assumed.
\section{Conlusion}
We investigated the influence of heat assistance in magnetic recording. Especially the modeling of an additional write assistance by using grains with high/low $T_{\mathrm{C}}$ layers was a main concern of this work. To deal with high temperatures in the vicinity of the Curie point we developed a computationally very cheap coarse grained LLB model. The LLB model treats each magnetic grain as a single magnetization vector. It requires detailed information about the temperature dependence of the zero field equilibrium magnetization $m_{\mathrm{e}}(T)$, the transverse and parallel susceptibilities $\widetilde{\chi}_{\perp}(T)$ and $\widetilde{\chi}_{\parallel}(T)$ and the intergrain exchange $A_{\mathrm{iex}}(T)$. In addition we derived an exact expression for the intergrain exchange field in the context of this LLB model. We proved that the LLB switching probabilities (under the influence of a Gaussian heat pulse and an external homogeneous magnetic field), of the coarse grained model fit the atomistic simulation results, obtained by the existing code VAMPIRE~\cite{evans_atomistic_2013} remarkably well, for strong as well as for weak intergrain exchange coupling. The speed-up of the LLB system compared to the atomistic calculations is formidable, which makes it easy to analyze the detailed influence of different heat pulses or other parameters with low computational effort, even for recording grains of realistic sizes. Additionally it would be possible to calculate the signal to noise ratio for a whole granular recording medium, which is presently out of reach for an atomistic code.\newline

The authors would like to thank the FWF Project SFB-ViCoM, F4112-N13 for the financial support. The support from the CD-laboratory AMSEN (financed by the Austrian Federal Ministry of Economy, Family and Youth, the National Foundation for Research, Technology and Development) and the support from Advanced Storage Technology Consortium (ASTC) is acknowledged. The computational results presented have been achieved using the Vienna Scientific Cluster (VSC).

\bibliography{LLB}

\end{document}